\documentclass[letterpaper]{article}
\usepackage[preprint]{aaai2027}
\usepackage[utf8]{inputenc}
\usepackage[hyphens]{url}
\usepackage{graphicx}
\usepackage{natbib}
\usepackage{caption}
\usepackage{amsmath,amssymb}
\usepackage{booktabs,tabularx,colortbl}
\usepackage{xspace}
\usepackage{algorithm}
\usepackage{algpseudocode}
\newcommand{\method}{\textsc{S1-MAS}\xspace}
\newcommand{\Description}[1]{}
\title{Token-Efficient Multi-Agent Collaboration via System One-Guided Computational Division of Labor}
\author{
    Zihan Zhou\textsuperscript{\rm 1},
    Xinzhe Hu\textsuperscript{\rm 1},
    Hanxu Yang\textsuperscript{\rm 1},\\
    Liangjian Wen\textsuperscript{\rm 2},
    Zhao Kang\textsuperscript{\rm 1,*}
}
\affiliations{
    \textsuperscript{\rm 1}University of Electronic Science and Technology of China\\
    \textsuperscript{\rm 2}Southwestern University of Finance and Economics\\
    2024080905013@std.uestc.edu.cn, zkang@uestc.edu.cn\\
    \textsuperscript{*}Corresponding author
}
\begin{document}
\maketitle
\begin{abstract}
Large language model (LLM)-based multi-agent systems (MAS) have become a promising paradigm for complex information-seeking and reasoning tasks by enabling collaborative problem solving among specialized agents. However, existing MAS frameworks tightly couple task reasoning with coordination operations, including task selection, role assignment, message routing, and context management. As interactions grow, using powerful LLMs for these bounded control decisions introduces substantial token overhead and latency, limiting the scalability of agentic Web services. In this paper, we investigate whether coordination can be decoupled from expensive reasoning without compromising collaborative performance. We propose S1-MAS, a token-efficient multi-agent framework based on System One-guided computational division of labor. S1-MAS assigns bounded coordination decisions to lightweight System One models while reserving open-ended reasoning for capable LLM workers. Specifically, a lightweight controller selects inspection conditions, chooses subsequent tasks, and determines termination, while a compact reader retrieves condition-relevant evidence from authorized sources to support these decisions. Through a decision--evidence loop, selected tasks dynamically determine worker roles and source access, enabling adaptive collaboration without task-specific training. Extensive experiments on seven diverse benchmarks demonstrate that S1-MAS achieves superior accuracy while substantially reducing the inference cost. Across individual comparisons with AgentVerse, DyLAN, and SelfOrg on seven benchmarks, S1-MAS reduces GPT-4o token consumption by 44.9\%--97.2\% and measured end-to-end latency by 37.8\%--93.0\%. These results highlight its potential for scalable and cost-effective agentic Web applications.
\end{abstract}

\section{Introduction}
\begin{figure}[!t]
\centering
\includegraphics[width=\columnwidth]{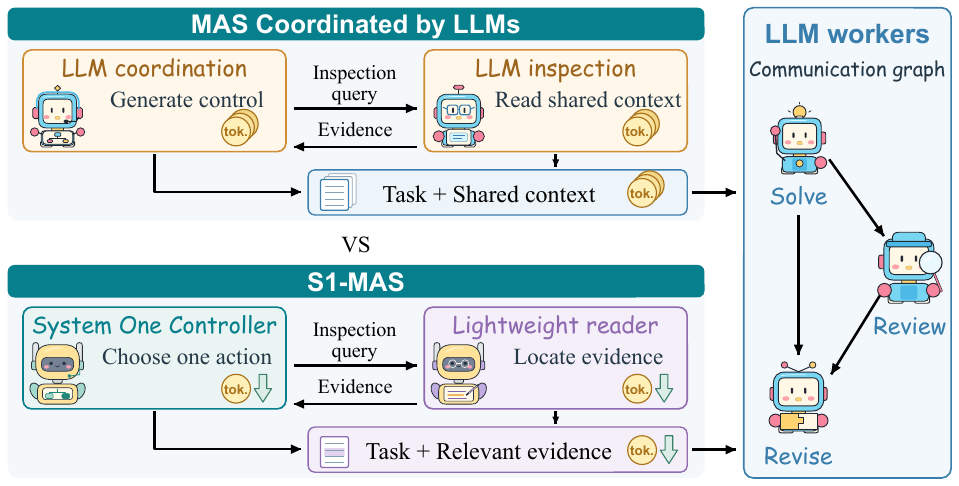}
\caption{S1-MAS decouples coordination from reasoning: a System One controller makes bounded coordination decisions,
while a lightweight reader localizes relevant evidence, reducing token overhead while preserving LLM reasoning capability.}
\Description{Two stacked rows contrast MAS coordinated by LLMs with S1-MAS. The upper row uses LLMs for coordination and inspection, with task and shared-context inputs passed to workers. Two downward Replace arrows lead to a System One controller that chooses one action and a lightweight reader that locates evidence. Paired arrows carry an inspection query and evidence between the two components. The lower row passes a task and relevant evidence to the same worker graph. Stacked token circles in the upper row and single circles with downward indicators in the lower row denote qualitative token reductions, not measured component-level ratios. The shared three-node graph contains Solve, Review, and Revise, with directed links from Solve to Review, Review to Revise, and Solve directly to Revise.}
\label{fig:orchestration}
\end{figure}

\begin{figure*}[t]
\centering
\includegraphics[width=\textwidth]{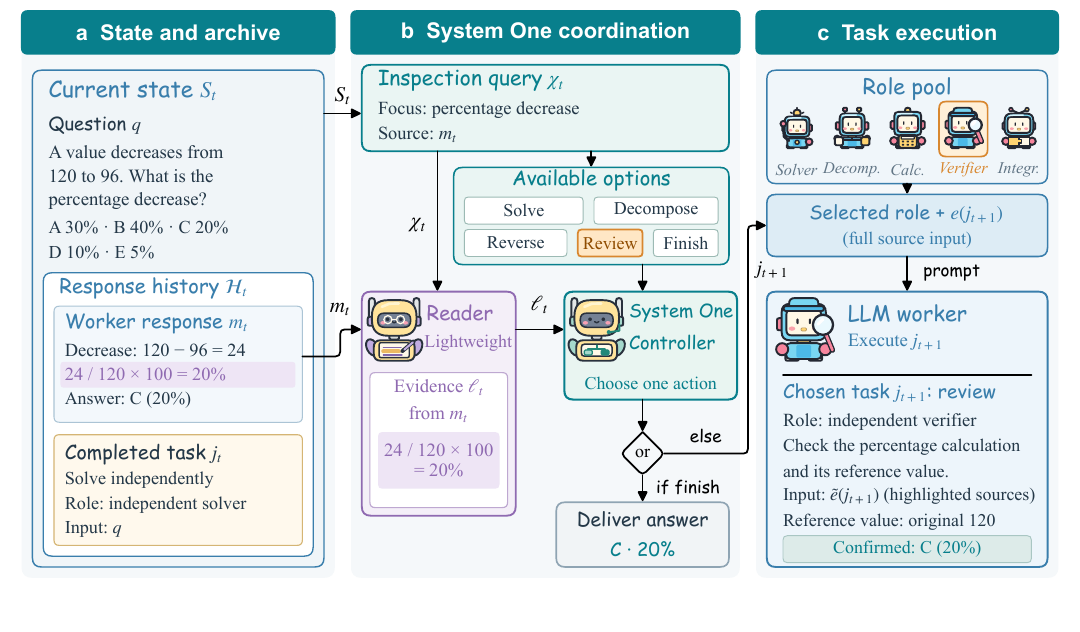}
\caption{Overview of S1-MAS.
(a) The state $S_t$ records the history of $t$ completed worker executions; in this AQuA example, the latest response $m_t$ generated by task $j_t$ serves as the current candidate answer.
(b) A lightweight System One controller makes bounded coordination decisions by selecting inspection conditions, choosing the next task, or terminating based on evidence localized by a lightweight reader.
(c) When collaboration continues, an LLM worker executes $j_{t+1}$ to generate $m_{t+1}$ and update the state to $S_{t+1}$; otherwise, the controller returns an archived answer upon termination.}
\Description{Three panels show state and archive, System One coordination, and task execution. Response history H sub t is nested inside current state S sub t, pairing worker response m sub t with completed task j sub t. The System One controller selects an inspection condition. A lightweight reader localizes condition-relevant evidence in the retained response. The Evidence ell sub t card is labeled from m sub t, with matching purple highlights indicating source correspondence rather than correctness. The displayed menu contains only the options eligible in this example; Repair and Contrast are unavailable in this state. The controller then chooses one action at a shared decision point. The else branch passes the selected task j sub t plus one to the LLM worker for execution; the mutually exclusive if-finish branch returns an archived answer. The role pool and complete authorized sources supply the worker prompt. The illustrated continuation is a review of the percentage calculation and its original reference value, 120.}
\label{fig:architecture}
\end{figure*}

Web-scale information services increasingly rely on large language model (LLM)-based agents to search, reason, and synthesize information from heterogeneous sources. Multi-agent systems (MAS) further extend this capability by allowing specialized agents to collaborate through iterative execution, evidence sharing, and response refinement~\cite{react,infogent,webwalker}. However, achieving effective collaboration requires not only high-quality reasoning but also continuous coordination: deciding which task should be executed, which agent should participate, what information should be accessed, and when the interaction should terminate. As agent trajectories become longer, the cost of coordination itself becomes a critical bottleneck for scalable agentic Web services.

Early MAS research studied task allocation, teamwork protocols, and communication strategies in distributed systems~\cite{contractnet,steam,commnet,dial,ic3net}. With the emergence of LLMs, agents can now perform flexible role-specific reasoning and coordinate through natural language interactions~\cite{autogen,agentverse}. Recent studies have explored adaptive orchestration mechanisms, including dynamic agent selection, communication topology construction, and workflow optimization~\cite{macnet,dylan,gdesigner,ofamas}. These methods improve collaboration flexibility; however, they generally rely on powerful generative LLMs to make coordination decisions. Consequently, the same models responsible for solving complex tasks are repeatedly invoked to generate control instructions, interpret intermediate histories, and maintain interaction states.

This design introduces a fundamental inefficiency. Many coordination operations in MAS are inherently bounded: selecting from feasible tasks, deciding whether additional inspection is required, or determining whether the current evidence is sufficient. Unlike open-ended reasoning, these operations do not require unrestricted generation. Treating them as general language generation wastes computation and creates unnecessary token consumption, especially when interaction histories continuously expand. Therefore, an important question arises:

\textit{Can coordination in multi-agent systems be performed by lightweight decision-making models while preserving the reasoning capability of powerful LLM workers?}

Recent advances in pretrained System One models provide a potential direction. Unlike generative LLMs, System One models are designed for fast and structured decision-making over constrained action spaces~\cite{jev,laya}. However, integrating them into MAS requires addressing a key challenge: coordination decisions must be grounded in relevant information from previous interactions rather than generated blindly. A lightweight controller needs compact and reliable evidence retrieval mechanisms that allow it to understand the current collaboration state without performing the entire reasoning process.

To address this challenge, we propose S1-MAS, a System One-guided framework for efficient multi-agent collaboration (Figure~\ref{fig:architecture}). S1-MAS introduces a computational division of labor between coordination and reasoning. A lightweight System One controller, instantiated with Laya~\cite{laya}, performs bounded coordination decisions, including selecting inspection conditions, assigning subsequent tasks, and determining termination. A lightweight reader, instantiated with Qwen3-0.6B~\cite{qwen3}, retrieves condition-relevant evidence from authorized sources and provides grounded information for controller decisions. Meanwhile, capable LLM workers remain responsible for open-ended reasoning, critique, and response refinement. Through this decision--evidence loop, S1-MAS enables adaptive collaboration while avoiding expensive LLM-based coordination generation.

We evaluate S1-MAS on seven diverse benchmarks covering mathematical reasoning, knowledge-intensive question answering, code generation, and Web-based question answering. Results show that S1-MAS achieves state-of-the-art accuracy among evaluated MAS baselines while substantially reducing computational cost. These results demonstrate that separating coordination from reasoning provides an effective pathway toward scalable and cost-efficient agentic Web services.

Our contributions are summarized as follows:

\begin{itemize}
\item \textbf{A new perspective on MAS efficiency.}
We identify coordination generation as a major computational bottleneck in LLM-based MAS and formulate a computational division of labor that separates bounded coordination from open-ended reasoning.

\item \textbf{System One-guided MAS framework.}
We introduce S1-MAS, the first systematic exploration of pretrained System One models for joint task orchestration and evidence-aware communication in general-purpose LLM-based MAS. S1-MAS integrates a lightweight controller, evidence reader, and LLM workers into a closed decision--evidence loop, enabling adaptive collaboration without task-specific training.

\item \textbf{Comprehensive evaluation.}
Experiments on seven benchmarks demonstrate that S1-MAS achieves competitive or superior accuracy. Across individual comparisons with AgentVerse, DyLAN, and SelfOrg on these benchmarks, it reduces GPT-4o token consumption by 44.9\%--97.2\% and measured end-to-end latency by 37.8\%--93.0\%.
\end{itemize}

\section{Related Work}
\textbf{Orchestration in Multi-Agent Systems.}
Organizing LLM agents requires not only assigning roles but also adapting collaboration patterns according to evolving task states. AgentVerse recruits specialized roles for collaborative problem solving~\cite{agentverse}, while MACNET explicitly models communication structures through directed acyclic graphs~\cite{macnet}. To enable adaptive collaboration during execution, DyLAN dynamically adjusts agent participation~\cite{dylan}, and SelfOrg reorganizes communication based on estimated peer contributions~\cite{selforg}. Beyond predefined strategies, recent approaches learn task-conditioned collaboration graphs~\cite{gdesigner,ofamas} or optimize next-agent and context selection~\cite{anymac}. These methods improve the flexibility of agent organization, but they still rely on expensive LLM-based decisions to determine how collaboration should proceed. S1-MAS addresses this complementary efficiency challenge by introducing a pretrained System One controller that separates bounded coordination decisions from LLM execution, enabling adaptive orchestration without task-specific training.

\textbf{Communication Optimization in Multi-Agent Systems.}
Efficient communication in MAS requires both reducing unnecessary exchanges and controlling the information involved in each interaction. AgentPrune addresses the structural aspect by pruning redundant communication edges~\cite{agentprune}. At the content level, MAD-M$^2$ masks debate memories~\cite{madmm}, while OPTiMACS learns compact message representations~\cite{optimacs}. Related prompt-compression approaches, including LLMLingua and LLMLingua-2, further reduce the context provided to language models~\cite{llmlingua,llmlingua2}. However, these approaches primarily optimize the communication payload after collaboration decisions are made, while the process of deciding what information to access or exchange can itself incur substantial LLM inference costs. \method addresses this complementary challenge by assigning source selection and evidence localization to lightweight components, reducing communication-control overhead without additional worker-LLM calls.

\textbf{Combining System One Models with LLMs.}
System One models, such as Jev and Laya, provide bounded and typed decisions without requiring free-form control generation~\cite{jev,laya}. Recent studies have explored their integration with LLM-based systems for action selection in LLM planning~\cite{jevmobile,jevstar}, edge-service orchestration~\cite{jevedge}, and agentic memory management~\cite{jevmem}. Security-oriented applications further investigate their use for agent pruning, finding adjudication, and confirmation loops~\cite{jevsecurity}. However, existing efforts primarily address control decisions within specific applications, leaving joint coordination and information exchange among LLM agents unexplored enough. Motivated by these advances, we propose \method, which is the first end-to-end LLM-based MAS framework that integrates pretrained System One models for joint task orchestration and evidence-based communication. Experiments demonstrate that \method substantially reduces LLM worker token consumption and end-to-end solution latency while maintaining collaborative task performance.

\section{Problem Formulation}
\label{sec:formulation}
Given a question $q$, we model multi-agent coordination as sequential decisions about which tasks workers execute, what information they receive, and when collaboration stops. Let $t$ denote the number of completed worker executions and $S_t$ the state of the system after these $t$ executions. The controller and reader calls are not counted in $t$.

\textbf{Communication graph.}
We use a directed acyclic graph to record which earlier responses are provided to subsequent worker executions:
\begin{equation}
 \mathcal G_t=(\mathcal V_t,\mathcal E_t),\qquad
 \mathcal V_t=\{v_i:1\le i\le t\}.
 \label{eq:graph}
\end{equation}
Here, $\mathcal V_t$ is the vertex set that represents completed worker executions and $\mathcal E_t$ is the directed edge set that represents direct response dependencies among these executions. Each vertex $v_i$ represents the execution $i$, where a worker performs the task $j_i$ under its assigned role and produces a response $m_i$. For $i<k\le t$, there is a directed edge $(v_i,v_k)\in\mathcal E_t$ if and only if the complete response $m_i$ is provided as a direct input to task $j_k$.

\textbf{State and archive.}
To guide subsequent coordination, we maintain an ordered archive
$\mathcal H_t$ of completed task--response pairs $(j_i,m_i)$, along
with the associated roles, sources, and observable check results.
The archive also includes an output identifier for comparing outputs
across responses: an option label, a short-text answer, or a piece of usable code, depending on the
answer format required by $q$.

Furthermore, we maintain an inspection ledger $\mathcal A_t$ to help avoid
repeated checks of whether a given response satisfies a particular
condition in $q$.
Specifically, $\mathcal A_t(i)$ stores records of attempts, up to
step $t$, to check whether the response $m_i$ satisfies the conditions
stated in $q$. These records track inspection attempts rather
than establish correctness.
We also track the remaining allowance $B_{\mathrm{src}}-u_t$
to constrain further transmission of source inputs to workers.
Here, $B_{\mathrm{src}}$ denotes the cumulative token budget for
source transmission, and $u_t$ denotes the amount consumed during
the executions of $t$ workers.
We combine this inspection history and the remaining allowance with
 question $q$ and archive $\mathcal H_t$ to define the current
state of coordination decisions.
\begin{equation}
 S_t=(q,\mathcal H_t,\mathcal A_t,B_{\mathrm{src}}-u_t).
 \label{eq:state}
\end{equation}

\textbf{Inspection and task selection.}
To provide explicit targets for inspecting a candidate answer, we partition the original question $q$ into sentences to form the set $\mathcal C(q)$. In state $S_t$, the System One controller forms an inspection query $\chi_t$ by pairing a selected sentence of $\mathcal C(q)$ with the current candidate answer. The query therefore specifies both what should be checked and which answer should be examined. The reader then locates the relevant text for this inspection target in the corresponding response, producing an evidence record $\ell_t$ for the controller's next decision.

Alongside the localized evidence, we provide the controller with
a menu of feasible tasks to select the next worker task.
Each candidate specifies an operation, a goal, a rule for assigning
a worker role, and the source responses needed for execution.
To determine which candidates enter this menu, a deterministic
compiler uses $S_t$ and $\chi_t$ to retain only those whose
prerequisites are met and whose required source inputs fit within
the remaining communication budget:
\begin{equation}
 \mathcal W_t=\operatorname{Compile}(S_t,\chi_t).
 \label{eq:compile}
\end{equation}
The set $\mathcal W_t$ contains feasible worker tasks; \texttt{finish} is considered separately when stopping is eligible. If collaboration continues, the controller uses $S_t$, $\chi_t$, and $\ell_t$ to select $j_{t+1}\in\mathcal W_t$. For the selected task, the set of sources satisfies $\mathcal P(j_{t+1})\subseteq\mathcal V_t$, and the input from the source $e(j_{t+1})$ contains the complete responses $\{m_i:v_i\in\mathcal P(j_{t+1})\}$ together. A worker assigned the compatible role executes $j_{t+1}$ and produces $m_{t+1}$, after which the system updates the state and communication graph to $S_{t+1}$ and $\mathcal G_{t+1}$. If the controller instead selects \texttt{finish}, collaboration stops and the deterministic delivery rule returns an archived response without another worker execution.

\section{Methodology}
\label{sec:framework}
S1-MAS assigns System One the role of a lightweight coordination controller: it selects what to inspect, which task to run, and when to stop. A lightweight reader localizes condition-relevant evidence, while LLM workers perform substantive reasoning. Laya and Qwen3-0.6B instantiate the controller and the reader, respectively. Their decision--evidence loop enables adaptive collaboration without requiring LLM workers to generate coordination instructions (Figure~\ref{fig:architecture}). We describe evidence localization (Section~\ref{sec:evidence}), task construction and selection (Section~\ref{sec:systemone}), role and source assignment followed by worker execution (Section~\ref{sec:communication}), and the adaptive stopping strategy (Section~\ref{sec:execution}). Algorithm~\ref{alg:uic} summarizes the overall procedure. All components operate without task-specific training.

\subsection{Inspection Query and Evidence Localization}
\label{sec:evidence}
The cumulative responses contain both useful evidence and details that are not related to the next coordination decision. Asking an LLM to repeatedly interpret this history leads to additional token consumption. Instead, we separate the choice of what to inspect from the location where it is addressed. Specifically, the System One controller selects a question condition to inspect, while the lightweight reader locates evidence relevant to that condition in the current candidate answer.

\textbf{System One controller: inspection query.}
In state $S_t$, we consider only the inspection targets in $\mathcal C(q)$ that have not yet been recorded for the current candidate answer in $\mathcal A_t$. The inspection is skipped when no candidate or eligible target is available.

To choose among the remaining targets, the controller compares them in small groups. Within each group, it evaluates the targets in forward and reverse menu orders and averages scores aligned by target identity to reduce sensitivity to presentation order. Selected targets advance through successive groups until one inspection target is retained or the controller chooses \texttt{skip}. The selected condition and the complete worker response corresponding
to the current candidate answer together form the inspection query
$\chi_t$.

\textbf{Lightweight reader: evidence localization.}
Given the question $q$ and the inspection query $\chi_t$, we construct
the evidence record as
\begin{equation}
 \ell_t=\mathcal{R}(q,\chi_t).
 \label{eq:evidence-localization}
\end{equation}
Here, $\mathcal{R}$ denotes the following
localization procedure. The lightweight reader selects identifiers
of source units relevant to the inspection condition within the full
worker response specified by $\chi_t$. We map these identifiers to
character offsets to retrieve the corresponding passages verbatim,
then store the passages, their offsets, the inspection condition,
and the source identity in $\ell_t$.

The controller may use $\ell_t$ as an additional context when selecting
the next action. When a worker's assigned task permits access to the
corresponding response, the localized passages can also serve as
attention cues. The record identifies material to examine rather
than certifying its correctness; substantive review remains the
worker's responsibility.

\subsection{Task Construction and System One-Guided Selection}
\label{sec:systemone}
To determine the next step of collaboration, the controller must
 select both the work to be performed and the information required
for its execution. We therefore use tasks as a unit of coordination. Each task
specifies an operation, a goal, a rule for assigning the worker
role, and the source responses the worker may use. A deterministic
compiler first identifies tasks whose prerequisites are met and
whose source inputs fit within the remaining communication budget.
The System One controller then uses the current state and available
evidence to choose among these feasible tasks. This separation makes
task selection a bounded decision over executable alternatives.

\textbf{Feasible task construction.}
A deterministic compiler builds the feasible menu $\mathcal W_t$ from $S_t$ and the inspection query $\chi_t$. The operations differ in their reasoning objectives and source requirements:
\begin{itemize}
\item \textbf{Solve:} Answer the original question independently, without using earlier responses.

\item \textbf{Decompose:} Divide the question into subproblems, solve them, and combine the results within one worker, without using earlier responses.

\item \textbf{Reverse:} Reason backward from the required outcome using the original question, without using earlier responses.

\item \textbf{Review:} Check the full worker response corresponding to the current candidate answer against the selected inspection target, revising it when concrete violations are identified. 
\item \textbf{Repair:} Correct missing answers, truncated responses, or code interface defects using the entire affected response. If all options are rejected without a parsed answer, solve the question again using only the original question.

\item \textbf{Contrast:} Compare the current candidate answer with an alternative response from the archive, using both responses in full. The alternative is the most recent response whose output identifier differs from the candidate's. 
\end{itemize}

These operations form a task vocabulary, and the feasible menu \(\mathcal W_t\) contains only currently admissible tasks and does not need to include all types of operation. The controller also considers \(\texttt{finish}\) when voluntary stopping is eligible.

\textbf{System One-guided action selection.}
The System One controller uses the current state $S_t$, inspection query $\chi_t$, and evidence record $\ell_t$ to select a task $j_{t+1}\in\mathcal W_t$ or the eligible \texttt{finish} option. In the next execution, a worker performs the selected task in its assigned role, while \texttt{finish} ends the collaboration and returns an archived answer without progress $t$.

\subsection{Task Execution and State Update}
\label{sec:communication}
\textbf{Role and assignment of sources.}
The selected task $j_{t+1}\in\mathcal W_t$ determines the role of the worker and the sources of responses that it receives. For example, selecting Review assigns the verifier role to the next worker. The worker receives the full response corresponding to the current candidate answer and checks it against the selected inspection target.

\textbf{Localized evidence as an attention cue.}
For a selected Review or Contrast task $j_{t+1}$, the evidence record $\ell_t$ identifies the relevant parts of an authorized source response to highlight, while the complete response is retained. Its source input with any applicable highlights is $\tilde e(j_{t+1})$; this retains all $e(j_{t+1})$ and equals $e(j_{t+1})$ when no cue is applied. This design narrows attention while retaining the context needed to challenge the highlighted reasoning.

\textbf{Execution of tasks by LLM workers.}
The worker receives the full question, the selected task, a compatible role, authorized sources, and an output contract specifying the required answer format or code interface. Using these inputs, it performs the corresponding operation: Solve, Decompose, Reverse, Review, Repair, or Contrast. Let $\operatorname{Prompt}$ construct this role-conditioned input, $\operatorname{LLM}$ denote worker generation, and $B_{\mathrm{out}}$ be the maximum number of tokens a worker can generate in each execution. Execution is
\begin{equation}
 m_{t+1}=\operatorname{LLM}
 \bigl(\operatorname{Prompt}(q,j_{t+1},\tilde e(j_{t+1}));B_{\mathrm{out}}\bigr).
 \label{eq:worker-execution}
\end{equation}

\textbf{Controlling communication budget.}
To save tokens, we limit the total number of source tokens transmitted across worker executions. All transmitted source tokens, including labels and attention cues, contribute to the following:
\begin{equation}
 u_t=\sum_{i=1}^{t}\operatorname{tok}(\tilde e(j_i))
 \le B_{\mathrm{src}}.
 \label{eq:source-budget}
\end{equation}
Here, $\operatorname{tok}$ counts the tokens in the transmitted source input. Copying the same response again consumes the budget repeatedly; tasks without earlier responses as input contribute zero. The communication budget restricts the repeated transmission of prior responses and the resulting context-processing overhead for LLM workers. By applying this limit across the entire workflow, S1-MAS keeps cumulative communication costs under control.

\textbf{Updating the interaction archive and the state of the system.}
Each response $m_{t+1}$ is archived with its completed task $j_{t+1}$, role, sources, output identifier, and observable checks. The execution of task $j_{t+1}$ adds a new vertex $v_{t+1}$ and incoming edges from its designated source vertices. The communication graph is updated as
\begin{equation}
 \begin{aligned}
  \mathcal V_{t+1} &= \mathcal V_t\cup\{v_{t+1}\},\\
  \mathcal E_{t+1} &= \mathcal E_t\cup
  \{(v_i,v_{t+1})\mid v_i\in\mathcal P(j_{t+1})\}.
 \end{aligned}
 \label{eq:edges}
\end{equation}
When $j_{t+1}$ performs Review or Contrast using $\chi_t$, the ledger records the attempted check for the response specified as the query's source. The new response $m_{t+1}$ inherits attempted-check records from a source response $m_i$ only if $v_i\in\mathcal P(j_{t+1})$ and the two responses share the same nonempty output identifier. Recording attempted checks avoids repeating the same review without treating the inspected content as verified. Updating $\mathcal H_{t+1}$, $\mathcal A_{t+1}$, and $u_{t+1}$ produces $S_{t+1}$; the delivery rule then identifies the current candidate answer from the updated archive.

\subsection{Adaptive Stopping and Answer Delivery Rule}
\label{sec:execution}
We formulate the stopping rule as
a bounded coordination decision: when stopping is eligible, the
System One controller can select \texttt{finish} instead of another
task. Once collaboration ends, a deterministic delivery rule returns
an existing response from the archive. This separates the decision
to continue computation from answer delivery, avoiding an additional
LLM call to evaluate or generate the final answer and the associated
token consumption.

\textbf{Adaptive stopping rules.}
Once the candidate answer exists, \texttt{finish} becomes eligible. The worker execution ends with the eligible \texttt{finish} option, selected by the controller or its fallback rule, or when the maximum number of worker executions $T_{\max}$ is reached. The execution limit is an enforced stop, not an additional controller decision.

\textbf{Answer delivery rules.}
In state $S_t$ with at least one completed execution, for multiple-choice questions and questions that require a numerical answer, the latest response $m_t$ takes precedence if it has a parsed answer key and task $j_t$ was an explicit two-source Contrast. Otherwise, delivery selects the most frequent nonempty answer key and its earliest supporting response, breaking the ties by the earliest occurrence. After each worker executes, this rule updates the current candidate answer. Subsequent inspection targets this candidate, which is also returned as the final output if collaboration stops.

For code generation tasks, delivery prioritizes the continuity of a usable implementation rather than the number of responses with the same answer. The first candidate that parses and provides the required function signature is retained. A later usable Review, Repair, or Contrast replaces it only if that task reads the retained candidate. If none is usable, delivery returns the last response. These public interface checks do not establish hidden-test correctness. Both output types therefore use the same coordination loop, with delivery contracts that return the original work rather than generate a new answer.

\begin{algorithm}[H]
\caption{S1-MAS: lightweight coordination and task execution}
\label{alg:uic}
\small\raggedright
\begin{algorithmic}[1]
\Require Complete question $q$, output contract, budgets $B_{\mathrm{src}},B_{\mathrm{out}}$, execution limit $T_{\max}$
\Ensure Current candidate answer from the archive
\State Initialize $S_0,\mathcal G_0$; $t,u_0\gets0$
\While{$t<T_{\max}$}
    \Statex \hspace{\algorithmicindent}\textbf{Inspection and localization}
    \State $\chi_t\gets\operatorname{SystemOne}(S_t,\mathcal C(q))$; $\ell_t\gets\varnothing$
    \State \textbf{if} $\chi_t\ne\varnothing$: $\ell_t\gets\mathcal{R}(q,\chi_t)$

    \Statex \hspace{\algorithmicindent}\textbf{Task selection}
    \State $\mathcal W_t\gets\operatorname{Compile}(S_t,\chi_t)$
    \State Controller scores $\mathcal W_t$ and eligible \texttt{finish} using $(S_t,\chi_t,\ell_t)$
    \State Select the highest-scoring option
    \State \textbf{if} \texttt{finish} is selected \textbf{then break}
    \State Otherwise, denote the selected task by $j_{t+1}\in\mathcal W_t$

    \Statex \hspace{\algorithmicindent}\textbf{Execution and state update}
    \State Resolve the selected task's role; form $\tilde e(j_{t+1})$
    \State $m_{t+1}\gets\operatorname{LLM}\bigl(\operatorname{Prompt}(q,j_{t+1},\tilde e(j_{t+1}));B_{\mathrm{out}}\bigr)$
    \State $u_{t+1}\gets u_t+\operatorname{tok}(\tilde e(j_{t+1}))$
    \State $\mathcal V_{t+1}\gets\mathcal V_t\cup\{v_{t+1}\}$
    \State $\mathcal E_{t+1}\gets\mathcal E_t\cup\{(v_i,v_{t+1}):v_i\in\mathcal P(j_{t+1})\}$
    \State Update $S_{t+1}$ (archive, ledger, budget)
    \State $t\gets t+1$
\EndWhile
\State \Return current candidate answer \Comment{No extra LLM call}
\end{algorithmic}
\end{algorithm}

\section{Experiments}
We evaluate S1-MAS to examine whether System One-guided coordination improves the efficiency of multi-agent collaboration while maintaining task performance. We first compare S1-MAS with representative baselines on six benchmarks covering reasoning and code generation, measuring task performance, GPT token consumption, and end-to-end latency. We then use ablation studies to examine how the System One controller and evidence localization contribute to task performance and inference efficiency. Finally, we use WebSRC to assess its effectiveness in answering questions on structured webpages.

\begin{table*}[t]
\caption{Performance comparison using GPT-4o. Accuracy values are reported as percentages. Results are means with standard deviations over three seeds. Average summarizes performance across six datasets. Best and second-best means are \textbf{bold} and \underline{underlined}, respectively.}
\label{tab:main}
\centering
\begingroup
\fontsize{9}{11}\selectfont
\setlength{\tabcolsep}{4pt}
\renewcommand{\arraystretch}{1.03}
\arrayrulecolor{black!75}
\begin{tabularx}{\textwidth}{>{\raggedright\arraybackslash}p{62pt}|*{6}{>{\centering\arraybackslash}X}|>{\centering\arraybackslash}X}
\toprule[0.8pt]
\multicolumn{1}{c|}{\textbf{Method}} & \textbf{MMLU} & \textbf{GSM8K} & \textbf{AQuA} & \textbf{MultiArith} & \textbf{SVAMP} & \textbf{HumanEval} & \textbf{Average}\\
\midrule[0.5pt]
\rowcolor{black!10}\multicolumn{8}{c}{\textbf{Direct Answer}}\\
Vanilla & 85.19{\scriptsize\,$\pm$1.00} & 94.93{\scriptsize\,$\pm$1.33} & 75.86{\scriptsize\,$\pm$4.34} & 98.81{\scriptsize\,$\pm$1.36} & 93.72{\scriptsize\,$\pm$0.26} & 85.67{\scriptsize\,$\pm$3.44} & 89.03{\scriptsize\,$\pm$1.26}\\
\specialrule{0.45pt}{2.5pt}{2.5pt}
\rowcolor{black!10}\multicolumn{8}{c}{\textbf{Single-Agent Prompting}}\\
CoT & 85.40{\scriptsize\,$\pm$1.64} & 97.13{\scriptsize\,$\pm$0.32} & 76.17{\scriptsize\,$\pm$1.87} & 99.23{\scriptsize\,$\pm$0.81} & 93.75{\scriptsize\,$\pm$0.81} & 87.60{\scriptsize\,$\pm$4.60} & 89.88{\scriptsize\,$\pm$0.97}\\
SC-CoT & 89.11{\scriptsize\,$\pm$1.00} & 97.30{\scriptsize\,$\pm$0.10} & 82.40{\scriptsize\,$\pm$3.32} & 99.40{\scriptsize\,$\pm$1.03} & 94.41{\scriptsize\,$\pm$1.73} & 88.98{\scriptsize\,$\pm$1.26} & 91.93{\scriptsize\,$\pm$0.57}\\
\specialrule{0.45pt}{2.5pt}{2.5pt}
\rowcolor{black!10}\multicolumn{8}{c}{\textbf{Fixed Multi-Agent Topologies}}\\
Chain & 88.24{\scriptsize\,$\pm$1.73} & 96.80{\scriptsize\,$\pm$0.10} & 80.37{\scriptsize\,$\pm$2.84} & 98.69{\scriptsize\,$\pm$0.74} & 94.31{\scriptsize\,$\pm$0.77} & 86.50{\scriptsize\,$\pm$2.66} & 90.82{\scriptsize\,$\pm$0.89}\\
Tree & 88.45{\scriptsize\,$\pm$1.36} & 95.63{\scriptsize\,$\pm$0.31} & 81.15{\scriptsize\,$\pm$1.89} & 99.40{\scriptsize\,$\pm$1.03} & 94.79{\scriptsize\,$\pm$0.31} & 88.71{\scriptsize\,$\pm$4.55} & 91.36{\scriptsize\,$\pm$0.96}\\
Complete & 88.67{\scriptsize\,$\pm$1.36} & 97.00{\scriptsize\,$\pm$0.70} & 82.24{\scriptsize\,$\pm$0.47} & 99.11{\scriptsize\,$\pm$0.78} & 95.21{\scriptsize\,$\pm$0.28} & 89.26{\scriptsize\,$\pm$0.83} & 91.91{\scriptsize\,$\pm$0.10}\\
Random & 87.80{\scriptsize\,$\pm$1.00} & 96.90{\scriptsize\,$\pm$0.79} & 80.06{\scriptsize\,$\pm$0.97} & 98.99{\scriptsize\,$\pm$0.88} & 94.65{\scriptsize\,$\pm$0.69} & 88.98{\scriptsize\,$\pm$1.72} & 91.23{\scriptsize\,$\pm$0.24}\\
\specialrule{0.45pt}{2.5pt}{2.5pt}
\rowcolor{black!10}\multicolumn{8}{c}{\textbf{Multi-Agent Coordination}}\\
MAD-M$^2$(S) & 88.89{\scriptsize\,$\pm$0.65} & 97.20{\scriptsize\,$\pm$0.70} & \underline{86.60}{\scriptsize\,$\pm$2.57} & 99.52{\scriptsize\,$\pm$0.82} & \underline{95.73}{\scriptsize\,$\pm$1.06} & 89.53{\scriptsize\,$\pm$1.26} & 92.91{\scriptsize\,$\pm$0.19}\\
SelfOrg & 88.45{\scriptsize\,$\pm$0.38} & 95.80{\scriptsize\,$\pm$0.26} & 81.31{\scriptsize\,$\pm$0.47} & 96.79{\scriptsize\,$\pm$0.18} & 92.67{\scriptsize\,$\pm$0.16} & 88.15{\scriptsize\,$\pm$1.26} & 90.53{\scriptsize\,$\pm$0.30}\\
AgentVerse & 90.20{\scriptsize\,$\pm$1.73} & 97.07{\scriptsize\,$\pm$0.72} & 84.11{\scriptsize\,$\pm$0.47} & 99.35{\scriptsize\,$\pm$1.13} & 95.17{\scriptsize\,$\pm$0.67} & 88.71{\scriptsize\,$\pm$3.91} & 92.43{\scriptsize\,$\pm$0.33}\\
MACNET & 89.11{\scriptsize\,$\pm$0.75} & 97.17{\scriptsize\,$\pm$0.91} & 82.55{\scriptsize\,$\pm$0.97} & 99.64{\scriptsize\,$\pm$0.62} & 95.03{\scriptsize\,$\pm$0.47} & \underline{90.36}{\scriptsize\,$\pm$2.52} & 92.31{\scriptsize\,$\pm$0.48}\\
DyLAN & \underline{91.07}{\scriptsize\,$\pm$1.00} & \underline{97.57}{\scriptsize\,$\pm$1.11} & 86.29{\scriptsize\,$\pm$2.31} & \underline{99.76}{\scriptsize\,$\pm$0.41} & 95.07{\scriptsize\,$\pm$0.93} & 88.98{\scriptsize\,$\pm$1.72} & \underline{93.12}{\scriptsize\,$\pm$0.99}\\
\specialrule{0.45pt}{2.5pt}{2.5pt}
\rowcolor{black!10}\multicolumn{8}{c}{\textbf{System One-Guided Orchestration}}\\
\textbf{S1-MAS} & \textbf{91.94}{\scriptsize\,$\pm$1.51} & \textbf{97.73}{\scriptsize\,$\pm$1.12} & \textbf{87.07}{\scriptsize\,$\pm$2.57} & \textbf{99.88}{\scriptsize\,$\pm$0.10} & \textbf{95.76}{\scriptsize\,$\pm$0.59} & \textbf{91.46}{\scriptsize\,$\pm$1.26} & \textbf{93.97}{\scriptsize\,$\pm$0.73}\\
\bottomrule[0.8pt]
\end{tabularx}
\endgroup
\end{table*}

\subsection{Experimental Setup}
\textbf{Datasets.}
We select seven benchmarks to evaluate S1-MAS in different reasoning requirements, output formats, and information sources. MMLU~\cite{mmlu} covers knowledge-based reasoning across subjects, while GSM8K~\cite{gsm8k}, AQuA~\cite{aqua}, MultiArith~\cite{multiarith}, and SVAMP~\cite{svamp} cover mathematical problem solving. HumanEval~\cite{humaneval} extends the evaluation from option labels and numerical answers to executable code. WebSRC~\cite{websrc} further assesses the answering of questions on structured websites, where agents must locate and use information from the website provided.

\textbf{Baselines.}
For the six main benchmarks, we compare S1-MAS with 12 baselines that do not require task-specific parameter training. Vanilla, CoT~\cite{cot}, and SC-CoT~\cite{selfconsistency} provide single-model references to assess the benefits of multi-agent collaboration. Chain, Tree, Complete, and Random represent collaboration under prescribed communication topologies, allowing comparison with adaptive coordination in S1-MAS. We further compare with five MAS frameworks -- AgentVerse~\cite{agentverse}, MACNET~\cite{macnet}, SelfOrg~\cite{selforg}, MAD-M$^2$(S)~\cite{madmm}, and DyLAN~\cite{dylan} -- to examine whether S1-MAS achieves a better balance between task performance and inference cost than existing approaches to agent orchestration and communication.

\textbf{Implementation details.}
We use GPT-4o for the six main benchmarks and WebSRC, and GPT-4o-mini for the ablation experiments. In WebSRC, we compare with AgentVerse, DyLAN, and SelfOrg, the three baselines evaluated on all seven benchmarks. Within each experiment, all methods use the same task-solving model, evaluation questions, response format requirements, and scoring procedures. Each LLM worker call uses a temperature of 1.0 and generates a single response, with at most $B_{\mathrm{out}}=2{,}048$ output tokens.

S1-MAS uses Laya as its System One controller and Qwen3-0.6B as its lightweight reader, without task-specific training. The reader uses greedy decoding in non-thinking mode. For each question, S1-MAS permits at most $T_{\max}=4$ worker executions, with dynamic stopping and no additional final judge. We set its cumulative budget for transmitting source responses to $B_{\mathrm{src}}=2{,}048$ tokens per question.

We repeat each comparison with three random seeds and report the mean accuracy. We also report the average GPT input and generation token counts and the average end-to-end latency per question. Token accounting includes every GPT call in the workflow. For the seven-benchmark comparison with AgentVerse, DyLAN, and SelfOrg, reduction ranges report the minimum and maximum over the 21 benchmark--baseline pairs.

\subsection{Main Results}
\label{sec:main-results}
\label{sec:efficiency}
\label{sec:latency}
We assess S1-MAS across six benchmarks in terms of task precision, GPT token consumption, and end-to-end latency.

\textbf{High performance.}
Table~\ref{tab:main} shows that S1-MAS achieves the highest mean accuracy on all six benchmarks. Its macro-average accuracy reaches 93.97\%, surpassing DyLAN and AgentVerse by 0.85 and 1.54 percentage points, respectively. The advantage spans different task demands: S1-MAS reaches 91.94\% on MMLU versus DyLAN's 91.07\%, and 87.07\% on AQuA versus MAD-M$^2$(S)'s 86.60\%. In HumanEval, it achieves 91.46\%, compared to 90.36\% for MACNET, the strongest alternative in that dataset. These results indicate that System One-guided collaboration supports high answer quality in knowledge, mathematical reasoning, and code generation.

\begin{table*}[t]
\caption{Mean GPT-4o tokens per question, averaged over three seeds under the same configurations as Table~\ref{tab:main} (lower is better). Ptok. and Ctok. denote input and generation tokens across all calls. Values are rounded to integers.}
\label{tab:tokens}
\centering
\begingroup
\fontsize{8}{10}\selectfont
\setlength{\tabcolsep}{3pt}
\renewcommand{\arraystretch}{1.12}
\arrayrulecolor{black!75}
\begin{tabularx}{\textwidth}{>{\raggedright\arraybackslash}p{64pt}|*{12}{>{\centering\arraybackslash}X}|*{2}{>{\centering\arraybackslash}X}}
\toprule[0.8pt]
\multicolumn{1}{c|}{\textbf{Method}} & \multicolumn{2}{c}{\textbf{MMLU}} & \multicolumn{2}{c}{\textbf{GSM8K}} & \multicolumn{2}{c}{\textbf{AQuA}} & \multicolumn{2}{c}{\textbf{MultiArith}} & \multicolumn{2}{c}{\textbf{SVAMP}} & \multicolumn{2}{c|}{\textbf{HumanEval}} & \multicolumn{2}{c}{\textbf{Average}}\\
& Ptok. & Ctok. & Ptok. & Ctok. & Ptok. & Ctok. & Ptok. & Ctok. & Ptok. & Ctok. & Ptok. & Ctok. & Ptok. & Ctok.\\
\midrule[0.5pt]
Vanilla & 117 & 25 & 80 & 98 & 105 & 176 & 62 & 48 & 62 & 39 & 161 & 155 & 98 & 90\\
MAD-M$^2$(S) & 4,066 & 1,407 & 2,907 & 1,235 & 4,764 & 2,045 & 2,164 & 729 & 2,052 & 697 & 6,418 & 2,734 & 3,728 & 1,475\\
SelfOrg & 9,795 & 5,429 & 7,032 & 3,254 & 10,910 & 5,979 & 4,682 & 1,922 & 4,835 & 2,097 & 17,513 & 10,909 & 9,128 & 4,932\\
AgentVerse & 7,018 & 3,562 & 6,780 & 3,205 & 9,645 & 5,206 & 4,156 & 1,688 & 4,332 & 1,826 & 14,070 & 7,728 & 7,667 & 3,869\\
MACNET & 2,356 & 1,165 & 2,329 & 1,270 & 3,472 & 2,097 & 1,511 & 722 & 1,539 & 745 & 3,984 & 2,272 & 2,532 & 1,379\\
DyLAN & 1,598 & 1,010 & 645 & 1,024 & 4,644 & 2,029 & 547 & 579 & 1,062 & 638 & 39,540 & 8,042 & 8,006 & 2,220\\
\midrule[0.4pt]
\textbf{S1-MAS} & \textbf{503} & \textbf{369} & \textbf{372} & \textbf{369} & \textbf{430} & \textbf{448} & \textbf{344} & \textbf{277} & \textbf{333} & \textbf{296} & \textbf{751} & \textbf{595} & \textbf{455} & \textbf{392}\\
\bottomrule[0.8pt]
\end{tabularx}
\endgroup
\end{table*}

\textbf{Token-efficient.}
Across the six main benchmarks, Table~\ref{tab:tokens} shows that S1-MAS reduces the mean consumption per-question of GPT tokens by 78.3--94.0\% relative to the five baselines of MAS, using fewer input and generation tokens in each benchmark. These savings accompany the accuracy gains rather than a trade-off in answer quality. System One makes coordination decisions without GPT-generated control instructions, while the lightweight reader provides relevant evidence. Task-dependent source access and a cumulative communication budget further limit repeated context transmission.

\begin{figure}[t]
\centering
\includegraphics[width=\columnwidth]{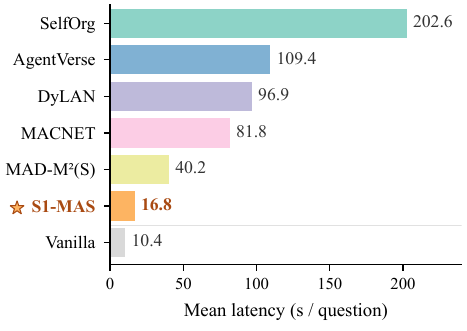}
\caption{Mean end-to-end latency across six benchmarks, averaged over three seeds under the same configurations as Tables~\ref{tab:main} and~\ref{tab:tokens}. Lower is better; the star highlights S1-MAS.}
\Description{Horizontal bars compare mean latency in seconds per question: SelfOrg 202.6, AgentVerse 109.4, DyLAN 96.9, MACNET 81.8, MAD-M2(S) 40.2, S1-MAS 16.8, and Vanilla 10.4. S1-MAS is highlighted with an orange star.}
\label{fig:latency}
\end{figure}

\textbf{Low-latency.}
Figure~\ref{fig:latency} shows that S1-MAS averages 16.8 seconds per question across the six main benchmarks, reducing the mean end-to-end latency by 58.2--91.7\% relative to the five MAS baselines, including the time spent on coordination and evidence localization. Beyond the lower coordination and context-processing overhead, adaptive stopping limits worker executions, and final delivery returns an archived response without an additional GPT call for judging or synthesis. These choices reduce sequential GPT generation, helping to explain the shorter solution time while reserving expensive calls for substantive reasoning.

\begin{table}[t]
\caption{Ablation results with GPT-4o-mini, averaged over three seeds. Accuracy (\%) is reported on MMLU, AQuA, and HumanEval. GPT tokens is the per-seed total over these three test sets, averaged across seeds.}
\label{tab:ablation}
\centering
\begingroup
\fontsize{8}{10}\selectfont
\setlength{\tabcolsep}{3pt}
\renewcommand{\arraystretch}{1.12}
\arrayrulecolor{black!75}
\begin{tabular*}{\columnwidth}{@{\extracolsep{\fill}}l|cccc|r@{}}
\toprule[0.8pt]
\textbf{Method} & \textbf{MMLU} & \textbf{AQuA} & \textbf{HumanEval} & \textbf{Average} & \shortstack{\textbf{GPT}\\\textbf{tokens}}\\
\midrule[0.5pt]
\textbf{S1-MAS} & \textbf{80.17} & \textbf{81.15} & \textbf{91.18} & \textbf{84.17} & 535,918\\
\midrule[0.4pt]
w/ Random & 78.43 & 78.50 & 90.08 & 82.34 & 542,491\\
w/ GPT & 78.87 & 79.91 & 90.63 & 83.14 & 2,190,250\\
w/o Reader & 79.08 & 78.35 & 89.81 & 82.41 & 482,314\\
\bottomrule[0.8pt]
\end{tabular*}
\endgroup
\end{table}

\subsection{Ablation Study}
\label{sec:ablation}
Table~\ref{tab:ablation} tests whether the accuracy--cost balance depends on how coordination and evidence localization are assigned. The variant w/ GPT delegates subsequent inspection and task-or-stop choices to GPT-4o-mini; w/ Random samples these choices uniformly from the legal menus. The w/o Reader variant removes evidence localization and its controller inputs and worker attention cues, but retains complete authorized parent responses. All variants share the worker model, role pool, evaluation settings, feasible-task rules, and budget limits.

\textbf{System One decisions improve collaboration.}
Replacing subsequent System One decisions with random routing reduces average accuracy from 84.17\% to 82.34\%, with decreases on all three datasets despite similar GPT token consumption. With the same task-construction rules and execution framework, this difference highlights the importance of selecting appropriate actions as the collaboration progresses. The System One controller selects inspection queries according to the current state, then uses the resulting evidence to choose the next task or stop. These decisions direct LLM workers toward useful follow-up work, enabling greater accuracy with comparable consumption of GPT tokens.

\textbf{System One makes better coordination decisions than the LLM controller.}
The complete S1-MAS achieves an average accuracy of 84.17\% using 0.54 million GPT tokens. In contrast, the GPT-routing variant consumes 2.19 million tokens but achieves a lower average accuracy of 83.14\%. S1-MAS therefore uses 75.5\% fewer GPT tokens in this controlled comparison. The assignment of subsequent coordination choices to the generative worker model substantially increases token use without improving mean task performance. This directly supports the central motivation for separating bounded decisions from open-ended LLM reasoning.

\textbf{The localization and highlighting of evidence improve the collaboration.}
Removing the reader reduces GPT token consumption, but lowers average accuracy. In S1-MAS, localized evidence informs subsequent task selection, while highlighting directs workers to relevant content within the source responses. Retaining these functions improves the average accuracy by 1.76 percentage points with 11.1\% more GPT tokens, demonstrating the value of evidence localization and highlighting for overall task performance.

\subsection{Effectiveness on Web-based Question Answering}
\label{sec:websrc}
To examine whether the benefits of S1-MAS extend to answering Web questions, we further evaluate it on WebSRC~\cite{websrc}, where answering a question requires locating relevant information within a structured webpage. All methods use GPT-4o and receive the same question and HTML of the webpage, without visual input. S1-MAS retains the coordination mechanism used in the main experiments. We compare it with AgentVerse, DyLAN, and SelfOrg, covering role-based collaboration, adaptive computation, and self-organized communication. GPT token accounting includes input and generation tokens for every call.

\begin{table}[t]
\caption{WebSRC performance and efficiency with GPT-4o. Exact match (EM) scores are averaged across three random seeds. GPT tokens and latency are per-question means over the same runs. Best results are bold.}
\label{tab:websrc}
\centering
\begingroup
\normalsize
\setlength{\tabcolsep}{4pt}
\renewcommand{\arraystretch}{1.12}
\arrayrulecolor{black!75}
\begin{tabular*}{\columnwidth}{@{\extracolsep{\fill}}lccc@{}}
\toprule[0.8pt]
\multicolumn{1}{c}{\textbf{Method}} & \textbf{EM (\%)} & \textbf{GPT tokens} & \textbf{Latency (s)}\\
\midrule[0.5pt]
AgentVerse & 89.78 & 32,386 & 28.78\\
DyLAN & 89.89 & 9,927 & 15.22\\
SelfOrg & 89.11 & 28,229 & 61.03\\
\midrule[0.4pt]
\textbf{S1-MAS} & \textbf{90.11} & \textbf{4,847} & \textbf{9.47}\\
\bottomrule[0.8pt]
\end{tabular*}
\endgroup
\end{table}

As shown in Table~\ref{tab:websrc}, S1-MAS achieves the highest mean exact-match accuracy and the lowest latency among the compared methods, while using 51.2--85.0\% fewer GPT-4o tokens than AgentVerse, DyLAN, and SelfOrg. These results indicate that the System One-guided framework remains effective when answering questions requires locating and using evidence from structured webpages.

\section{Conclusion}
S1-MAS introduces a System One-guided computational division of labor that decouples adaptive coordination from generative task reasoning. Its central insight is that deciding how agents collaborate and exchange information can be formulated as bounded choices informed by the current state and available evidence, rather than repeatedly generating coordination instructions with powerful LLMs. This formulation enables System One to organize adaptive collaboration without task-specific training while reserving the LLM computation for substantive reasoning. Comprehensive experiments demonstrate substantial reductions in GPT token consumption while maintaining strong task performance. Routing Ablation studies further show that System One achieves higher accuracy than random and GPT-based routing, establishing the value of its decisions beyond their low computational cost. Together, these findings highlight System One as an effective basis for coordinating agent execution and communication, offering a direction for efficient agentic Web systems. Future work can extend this approach to multi-page Web retrieval and question answering.

\bibliography{references}

\end{document}